\documentclass[final,5p,times,twocolumn]{elsarticle}
\usepackage{graphicx,amssymb,dcolumn,bm,amsmath,natbib}
\usepackage{indentfirst}
\usepackage{epsfig}
\usepackage{epstopdf}
\usepackage{exscale}
\usepackage[colorlinks=true,citecolor=black]{hyperref}
\usepackage{caption2}
\usepackage{relsize}

\journal{Chaos, Solitons \& Fractals}
\biboptions{sort&compress}
\biboptions{}

\begin{document}

\begin{frontmatter}

\title{Stable higher-order vortex quantum droplets in an annular potential}

\author{Liangwei Dong \corref{cor1}}
\address{Department of Physics, Zhejiang University of Science and Technology, Hangzhou, China, 310023}
\cortext[cor1]{Corresponding author.}
\ead{dlw0@163.com}

\author{Mingjing Fan}
\address{Department of Physics, Shaanxi University of Science $\&$ Technology, Xi'an, China, 710021}

\author{Boris A. Malomed}
\address{Instituto de Alta Investigacion, Universidad de Tarapaca, Casilla 7D, Arica, Chile** \corref{cor2}}
\cortext[cor2]{Sabbatical address.}
%
%
%

\date{\today}




\begin{abstract}
We address the existence, stability, and evolution of two-dimensional vortex quantum droplets (VQDs) in binary Bose-Einstein condensates trapped in a ring-shaped potential. The interplay of the Lee-Huang-Yang-amended nonlinearity and trapping potential supports two VQD branches, controlled by the radius, width and depth of the potential profile. While the lower-branch VQDs, bifurcating from the system's linear modes, are completely unstable, the upper branch is fully stable for all values of the topological charge $m$ and potential's parameters. Up to $m=12$ (at least), stable VQDs obey the {\it anti-Vakhitov-Kolokolov} criterion. In the limit of an extremely tight radial trap, the modulational instability of the quasi-1D azimuthal VQDs is studied analytically. We thus put forward an effective way to produce stable VQDs with higher vorticity but a relatively small number of atoms, which is favorable for experimental realization. 
\end{abstract}


\end{frontmatter}

\section{Introduction}

\label{Sec1}

The past few years have witnessed significant advancements in studies of the
evolution dynamics of quantum droplets (QDs) \cite%
{PhysRevLett.115.155302,PhysRevLett.117.100401,NATU2016,PhysRevLett.117.215301,Science2018,Phystoday2019}%
. QDs draw much interest due to their unique properties. In particular, the
quantum liquid of which the QDs are made is eight orders of magnitude more
dilute than liquid helium, constituting the most dilute liquid known in
physics \cite{NATU2016,PhysRevLett.117.215301,Science2018}. QDs behave as
nearly incompressible self-sustained liquid droplets, with the core
featuring a uniform density in the limit of large atom numbers \cite%
{PhysRevLett.115.155302}. QDs may be stable not only in their fundamental
state, but also in excited forms, e.g., as vortex QDs (VQDs) or
multipole-mode QDs in both two-dimensional ($2$D) \cite{PhysRevA.98.063602,
PhysRevA.106.053303, PhysRevA.105.033321} and three-dimensional ($3$D) \cite%
{PhysRevA.98.013612,DONG2023113728} geometries.

The possibility of creating stable QDs in binary Bose-Einstein condensates
(BECs) with contact inter-atomic interactions was theoretically predicted by
Petrov \emph{et al.} \cite{PhysRevLett.115.155302, PhysRevLett.117.100401},
assuming intrinsic self-repulsion in each component and dominating
attraction between them, imposed by means of the Feshbach resonance \cite{PhysRevLett.99.010403}. Further, the attraction-induced collapse is arrested by the self-repulsive
correction to the BEC\ energy resulting from zero-point quantum fluctuations
around the mean-field state. The latter was derived by Lee, Huang, and Yang
(LHY) in 1957 \cite{PhysRev.106.1135}, and extended to binary mixtures in
1963 \cite{LARSEN196389}. Experimentally, anisotropic QDs were created in
dipolar BECs by exploiting the competition between the contact repulsion and
dipole-dipole attraction \cite%
{Nature2016_3,PhysRevLett.116.215301,PhysRevX.6.041039,NATU2016,li2024two}. Quasi-2D
\cite{Science2018,PhysRevLett.120.135301} and full 3D isotropic QDs \cite%
{PhysRevLett.120.235301} were observed in mixtures of two different atomic
states in $^{39}$K and in an attractive mixture of $^{41}$K and $^{87}$Rb
atoms \cite{PhysRevResearch.1.033155}, which feature solely contact
interactions. QDs in binary dipolar BECs were also reported recently \cite%
{PhysRevLett.126.025301,PhysRevLett.126.025302}.

Theoretical works have confirmed close agreement between properties of QDs
predicted by the LHY-amended Gross-Pitaevskii equations (GPEs) and by the
many-body theory based on the quantum diffusion Monte-Carlo method \cite%
{PhysRevA.99.023618,NJP2020,PhysRevLett.122.105302,PhysRevLett.119.215302,PhysRevLett.117.205301, PhysRevB.97.140502}. Thus, properties of QDs were predicted in the framework of the LHY-amended
GPEs in various settings \cite{PhysRevA.98.063602, PhysRevA.106.053303,
PhysRevA.105.033321,PhysRevA.98.013631,ZHOU2019104881,CHEN2021103781,ZHOU2021111193, PhysRevResearch.2.033522,PhysRevLett.122.193902,PhysRevLett.123.160405,PhysRevLett.123.133901, zheng2021quantum,JIANG2022112368,HUANG2022112079,ZHAO2022112481, HUANG2023113137,XU2022112665,luo2021new}. In particular, it was predicted that 2D and 3D asymmetric QDs may perform circular motion in an anharmonic trapping potential \cite{DongPRL2021}.

Vortices are ubiquitous objects appearing in various branches of nonlinear
physics, including BECs, nonlinear optics, classical and quantum
hydrodynamics, magnetism, etc. \cite{pismen1999vortices}. They can be generated by means
of various experimental techniques and have found extensive applications in
a variety of fields, such as quantum computing, optical trapping, and
superconductors \cite{shen2019optical}. Specifically, while $2$D fundamental
QDs in symmetric binary BECs are completely stable, their stable VQDs
counterparts can be found with the topological charge up to $m=5$, provided
that the VQD's atom number (norm) exceeds a certain critical value \cite%
{PhysRevA.98.063602}. Stable $3$D VQDs have been found, thus far, with $m=1$
and $2$ \cite{PhysRevA.98.013612}. Internal modes of stable $2$D VQDs were
studied in Ref.~\cite{PhysRevA.106.053303} (see also Ref. \cite{PhysRevA.101.063628},
where results were reported for intrinsic stability of 2D bound states
stabilized by the LHY self-repulsion in the singular trapping potential $%
\sim -1/r^{2}$). Binary QDs with heterosymmetric structures in their
components were studied too \cite{PhysRevResearch.2.033522}. Metastable
globally-linked QD clusters and rotating VQD ones with multiple singly
quantized vortices were predicted in symmetric binary BECs \cite{PhysRevLett.122.193902,PhysRevLett.123.160405}. Semidiscrete optical vortex droplets can exist in quasi-phase-matched photonic crystals \cite{Xu:23}.

In the absence of external trapping, VQDs with high topological charges are
stable only for BECs containing an extremely large number of atoms \cite%
{PhysRevA.98.063602, PhysRevA.98.013612}. Then, challenging issues are if it
is possible to design a practically relevant setting for the realization of
stable VQDs with higher values of the topological charge (e.g., $m=12$), if
VQDs with large $m$ may be made stable with relatively small atom numbers,
and whether one can control the distribution of VQDs in such settings.

In this paper, we investigate the existence, stability, and dynamics of VQDs
in symmetric binary BECs trapped in an annular potential. The analysis
reveals that the radius, thickness and depth of the annular potential can be
used to control VQD properties. In particular, VQDs belonging to an upper
branch of solutions are stable in their entire existence domains. We demonstrate that stable VQDs with higher values of $m$ can be found for symmetric binary BECs containing relatively small numbers of atoms. In addition to the systematic numerical studies, analytical
consideration is performed for the modulational instability (MI) of
azimuthally uniform VQDs in the quasi-1D limit of an extremely narrow and
deep annular trap. These findings suggest the way for the experimental
creation of elusive self-sustained higher-charge $2$D vortex droplets in BECs.


\section{The theoretical model}

\label{Sec2} The evolution of wave functions $\Psi _{1,2}(x,y,z,t)$ of the
two components of the $3$D binary BECs is governed by the GPE system with
the mean-field cubic terms augmented by the LHY-induced quartic
self-repulsion \cite{PhysRevLett.115.155302}. For the BEC strongly confined
in the transverse direction, the full GPE system can be reduced to the $2$D
system for modes with lateral size $l\gg \sqrt{a_{1,2}a_{\bot }}$, where $%
a_{1,2}$ and $a_{\bot }$ are the self-repulsion scattering lengths of the
two component and the transverse-confinement length, respectively \cite%
{LiNJP2017}. Typical parameters relevant to the experiments with the
quasi-2D setups, \textit{viz}., $l\sim 10$ $\mathrm{\mu }\text{m},a_{{1,2}%
}\sim 3\text{ nm}$ and $a_{\bot }\lesssim 1$ $\mathrm{\mu }\text{m }$\cite%
{PhysRevLett.115.155302,PhysRevLett.117.100401}, satisfy this condition. In
the scaled form, the effective 2D system is \cite{PhysRevLett.117.100401}

\begin{equation}\label{Eq1}
  \begin{aligned}
  i\frac{\partial\Psi_{{1,2}}}{\partial t} =&-\frac{1}{2}\nabla^2\Psi_{1,2} +V(x,y)\Psi_{{1,2}}+\frac{4\pi}{g}\left(|\Psi_{1,2}|^2-|\Psi_{{2,1}}|^2\right)\Psi_{{1,2}} \\
  &+\left(|\Psi_1|^2+|\Psi_{2}|^2\right)\ln\left(|\Psi_1|^2+|\Psi_{2}|^2\right)\Psi_{{1,2}}.
  \end{aligned}
\end{equation}
where $\nabla ^{2}=\partial _{xx}+\partial _{yy}$, $V\left( x,y\right) $ is
the external potential, and $g>0$ is the coupling constant. Further, we
focus, as usual, on the symmetric system, with, $a_{1}=a_{2}$ and $\Psi
_{1}=\Psi _{2}=\Psi /\sqrt{2}$. With the addition rescaling, $%
(x,y)\rightarrow (g/2\sqrt{\pi })(x,y)$ and $t\rightarrow (g^{2}/4\pi )t$,
the final form of the 2D LHY-amended GPE becomes

\begin{equation}
i\frac{\partial \Psi }{\partial t}=-\frac{1}{2}\nabla ^{2}\Psi +|\Psi
|^{2}\ln (|\Psi |^{2})\Psi +V\left( x,y\right) \Psi .  \label{Eq2}
\end{equation}

We consider the QD dynamics under the action of an annular potential,
defined as
\begin{equation}
V(r)=-p\exp \left[ -(r-r_{0})^{2}/d^{2}\right] ,  \label{Eq3}
\end{equation}%
where $\left( r,\theta \right) $ are the polar coordinates, $r_{0}$, $d$,
and $p$ being the radius, width, and depth of the annular potential. The most relevant case is the one with the potential depth being on the same order of magnitude as the spatial scale corresponding to the nonlinear term in Eq.~(\ref{Eq2}).

Stationary solutions of Eq.~(\ref{Eq2}) with chemical potential $\mu $ are
searched for as $\Psi (x,y,t)=\psi (x,y)\exp (-i\mu t)\equiv \left[\psi
_{r}(x,y)+i\psi _{i}(x,y)\right] \exp (-i\mu t)$, where $\psi _{r}$ and $\psi _{i}$ are the real and imaginary parts of the stationary wave function,
whose phase is $\phi =\arctan \left( \psi _{i}/\psi _{r}\right) $. The
substitution of this ansatz in Eq.~(\ref{Eq2}) results in a stationary
equation
\begin{equation}
\frac{1}{2}\nabla ^{2}\psi +\mu \psi -V(r)\psi -|\psi |^{2}\ln (|\psi
|^{2})\psi =0,  \label{Eq4}
\end{equation}%
from which stationary states can be looked for by dint of the relaxation or
Newton-conjugate-gradient method \cite{Book2}. QD families are characterized
by the set of parameters $\mu $, $r_{0}$, $d$ and $p$.

Equation (\ref{Eq2}) conserves three dynamical invariants, namely, the norm
(scaled number of atoms) $N$, Hamiltonian (energy) $E$, and the total
angular momentum $M_{z}$:
\begin{equation}
\begin{aligned}
&N=\int\int|\Psi|^2\text{d}x\text{d}y, \\
&E=\frac{1}{2}\int\int\left[|\nabla\Psi|^2+2V|\Psi|^2+|\Psi|^4\ln\left(\frac{|\Psi|^2}{\sqrt{e}}\right) \right] \text{d} x \text{d} y, \label{Eq5} \\
&M_z=i\mathlarger{\int}\mathlarger{\int}\left[ \Psi^*\left(y \frac{\partial}{\partial x}-x \frac{\partial}{\partial y}\right)\Psi\right]\text{d}x\text{d}y \equiv i\mathlarger{\int}\mathlarger{\int}\Psi^*\frac{\partial }{\partial \theta}\Psi\text{d}x\text{d}y.
\end{aligned}
\end{equation}
For vortex states with topological charge $m$, $\Psi (x,y,t)=\psi (x,y)\exp
(im\theta -i\mu t)$, one has $M_{z}=mN$.

The analysis of the QD stability is crucially important for identifying
physically relevant states. To this end, one should use the linearized
Bogoliubov-de Gennes equations (BdGEs) \cite{Book11,Book22} for small
perturbations $f\left( x,y\right) $ and $g\left( x,y\right) $ of the wave
functions, defined so that
\begin{eqnarray}
\Psi (x,y,t)&=&[\psi +f\left( x,y\right) \exp (\delta t)  \notag \\
&&+g^{\ast }\left( x,y\right) \exp (\delta ^{\ast }t)]\exp (-i\mu t),
\label{Eq6}
\end{eqnarray}%
where $\delta$ is the (generally, complex) growth rate of the perturbation,
and $\ast$ stands for the complex conjugate. The linearization of Eq.~(\ref{Eq2}) for the small perturbations leads to the linear-stability eigenvalue
problem
\begin{equation}
\centering\delta \left[
\begin{array}{c}
f \\
g \\
\end{array}%
\right] =i\left[
\begin{array}{cc}
M_{1} & M_{2} \\
-M_{2}^{\ast } & -M_{1}^{\ast }
\end{array}%
\right] \left[
\begin{array}{c}
f \\
g \\
\end{array}%
\right] .  \label{Eq7}
\end{equation}%
Here, $M_{1}=-(1/2)\nabla ^{2}+V(r)-\mu +2|\psi |^{2}\left[ \ln (|\psi
|^{2})+1/2\right] $ and $M_{2}=\psi ^{2}\left[ \ln (|\psi |^{2})+1\right] $.
Equations (\ref{Eq7}) can be solved numerically by means of the Fourier
collocation algorithm \cite{Book2}. The QD stability is determined by the
spectrum of Eqs.~(\ref{Eq7}). QDs are stable only when all eigenvalues $%
\delta $ are pure imaginary.

\section{The quasi-one-dimensional limit (narrow annular trap): analytical
results}

Before presenting results of the systematic numerical investigation for the
VQDs, which are supported by the annular trapping potential in its general
form, and their stability, it is relevant to consider the limit case of a
very narrow and deep trap, which makes it possible to produce analytical
results. In this limit, the narrow radial Gaussian (\ref{Eq3}), with small $d
$ and large $p$, may be approximated as%
\begin{equation}
V(r)=p\exp \left[ -(r-r_{0})^{2}/d^{2}\right] \approx \varepsilon \delta
\left( r-r_{0}\right) ,\varepsilon \equiv \sqrt{\pi }pd\gg r_{0}^{-1},
\label{epsilon}
\end{equation}%
where $\delta \left( r-r_{0}\right) $ is the delta-function. This trapping
potential maintains bound states which are strongly confined in the radial
direction, \textit{viz}.,
\begin{equation}
\Psi \left( r,\theta ,t\right) \approx \sqrt{\varepsilon }\exp \left( -\frac{%
i}{2}\varepsilon ^{2}t-\varepsilon \left\vert r-r_{0}\right\vert \right)
\Phi \left( \theta ,t\right) ,  \label{Phi}
\end{equation}%
where the $r$-dependent factor is the normalized wave function of the
commonly known 1D bound state supported by the delta-functional trapping
potential. The reduced GPE for the azimuthal wave function, $\Phi \left(
\theta ,t\right) $, can be derived, as usual \cite{PhysRevA.65.043614}, by the
substitution of the factorized ansatz (\ref{Phi}) in the underlying equation
(\ref{Eq2}), multiplying it by the same radial wave function as is present
in the ansatz, and finally performing the radial integration. The results is%
\begin{equation}
i\frac{\partial \Phi }{\partial t}=-\frac{1}{2r_{0}^{2}}\frac{\partial
^{2}\Phi }{\partial \theta ^{2}}+\frac{\varepsilon }{2}\left\vert \Phi
\right\vert ^{2}\Phi \ln \left( e^{-1/8}\varepsilon \left\vert \Phi
\right\vert ^{2}\right) .  \label{azimuthal}
\end{equation}

It is relevant to stress that the effective 1D equation (\ref{azimuthal})
keeps the 2D form of the nonlinear term in the case when the
radial-localization range, $\sim 1/\varepsilon $ in Eq.~(\ref{Phi}) (which
is on the order of $\mathrm{\mu }$m, in experimentally relevant settings),
is much smaller than $r_{0}$, but large in comparison to the underlying
healing length, which is typically $\xi \sim 30$ nm \cite{Science2018,PhysRevLett.120.135301,PhysRevLett.120.235301,PhysRevResearch.1.033155}. In the limit of the extremely tight-confinement range, which is smaller
than the healing length, the LHY correction to the effective 1D GPE is
different, being represented by the self-attractive quadratic term, $\sim
\left\vert \Phi \right\vert \Phi $ \cite{PhysRevLett.117.100401}. In
particular, the MI of spatially uniform states, which is the central issue
of the analytical consideration presented in this section, was analyzed, in
the framework of the latter model, in Refs. \cite{sym12010174, TABI2023129087, otajonov2022modulational}.

Getting back to Eq.~(\ref{azimuthal}), which is the experimentally relevant
model in the present context, its azimuthally uniform solutions with
amplitude $A$, carrying integer vorticity $m$, are%
\begin{eqnarray}
\Phi &=&A\exp \left( -i\mu t+im\theta \right) ,  \notag \\
\mu &=&\frac{m^{2}}{2r_{0}^{2}}+\frac{\varepsilon }{2}A^{2}\ln \left(
e^{-1/8}\varepsilon A^{2}\right) .  \label{uniform}
\end{eqnarray}%
In terms of the quasi-1D limit, the instability of the higher-order VQD
states, with large values of $m$, corresponds to the MI of solutions (\ref%
{uniform}) in the framework of Eq.~(\ref{azimuthal}). The Galilean
invariance of this equation implies that the stability does not depend on $m$%
, which helps to understand the fact that the VQD stability, analyzed below
in terms of the full 2D model (\ref{Eq2}), readily extends to solutions with
large values of $m$.

Thus, the MI of solutions (\ref{uniform}) depends solely on their amplitude $%
A$. The straightforward analysis, which takes into regard the
\textquotedblleft quantization" of wavenumbers of the modulational
perturbations, which may assume only integer values, $p=\pm 1,\pm 2,...$,
leads to the following MI condition:%
\begin{equation}
A^{2}\ln \left( e^{7/8}\varepsilon A^{2}\right) <-p^{2}/\left(
2r_{0}^{2}\varepsilon \right) .  \label{MI}
\end{equation}%
Further, the analysis of condition (\ref{MI}) demonstrates that this
condition \emph{does not hold} for all $p^{2}\geq 1$, provided that the
radius of the narrow trapping potential is not too large, \textit{viz}.,
\begin{equation}
r_{0}\leq e^{15/16}/\sqrt{2}\approx 1.8.  \label{r0}
\end{equation}%
Thus, condition (\ref{r0}) guarantees the full stability of all vortex
states in the framework of the quasi-1D limit (\ref{azimuthal}).

\section{Numerical results}

\label{Sec3}
\begin{figure}[tbph]
\centering
\includegraphics[width=0.48\textwidth]{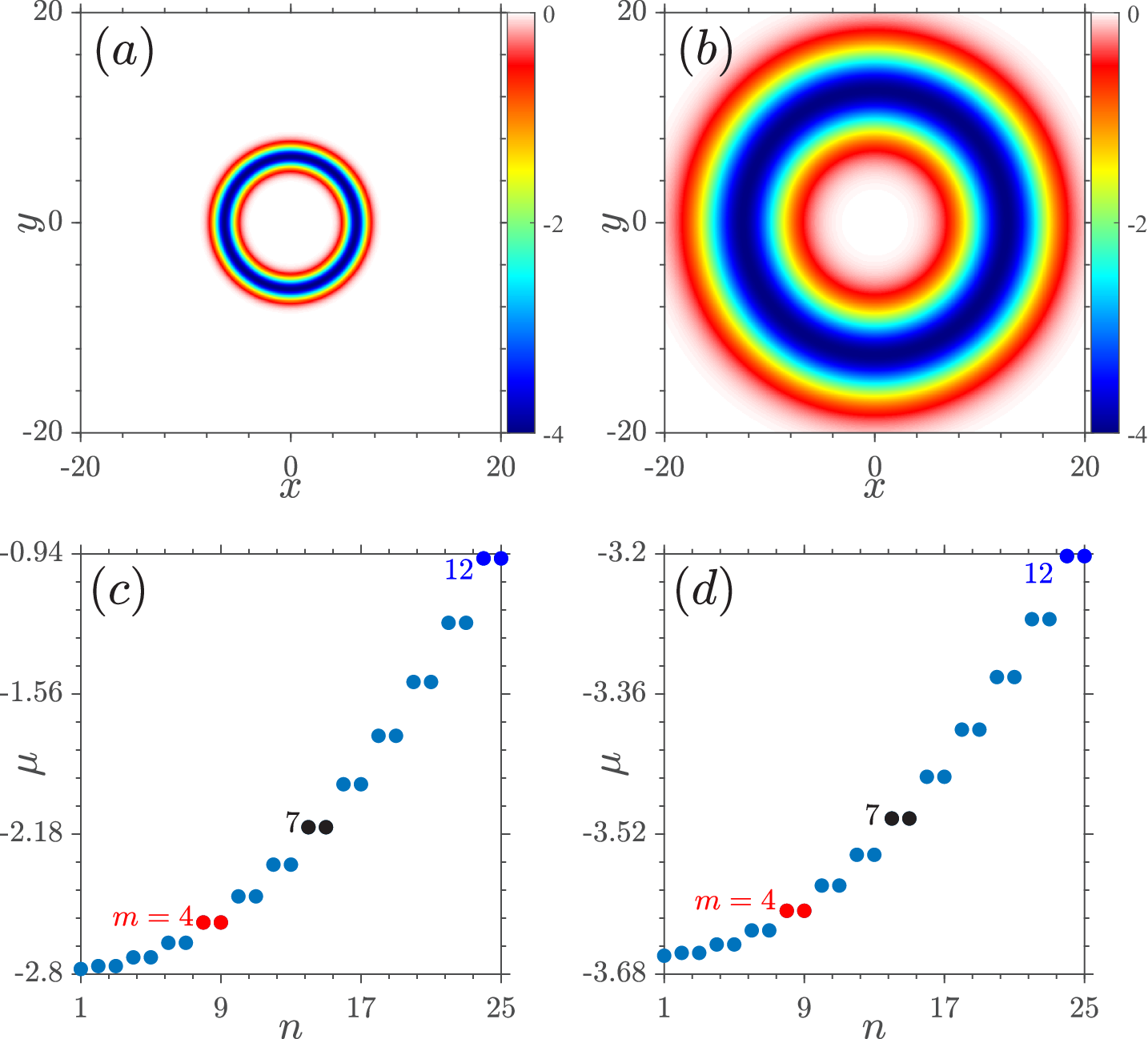}\vskip-0pc
\caption{Examples of the annular potential (\protect\ref{Eq3}) with $r_{0}=2%
\protect\pi, d=1$ (a) and $r_{0}=4\protect\pi ,d=4$ (b). (c,d) Spectra of
eigenvalues $\protect\mu $ of the linearized equation (\protect\ref{Eq4})
for the potentials shown in (a) and (b), respectively. Labels $m$ denotes
eigenvalues from which VQDs with topological charges $m=4,7$ and $12$
bifurcate. In all the panels, the potential depth is $p=4$. }
\label{fig1}
\end{figure}

To explore effects of the radius and width of the ring-shaped potential (\ref%
{Eq3})\ on the QDs, we fix the potential's depth as $p=4$ and vary its
radius $r_{0}$ and width $d$. Two representative examples of the potential
are presented in Figs.~\ref{fig1}(a,b). Note that the potential is symmetric
about the central circle with $r=r_{0}$, i.e., $V(r=r_{0}-r^{\prime
})=V(r=r_{0}+r^{\prime })$, for all $0<r^{\prime }<r_{0}$. This symmetry
implies the same guidance for the wave function in the inner and outer
annuli ($r<r_{0}$ and $r>r_{0}$, respectively), unlike the case of the
harmonic-oscillator (HO)\ trapping potential, which is commonly used to trap
BEC and guide optical beams. We have found that the potential defined by
Eq.~(\ref{Eq3}) suppresses the instability of vortex solitons with high
values of the topological charge which is known in the case of the HO
trapping potential \cite{Dong:23}. Stable higher-order solitons with a
multiring profile are also expected to be maintained by potential (\ref{Eq3}).

Before addressing the VQDs in the nonlinear regime, it is instructive to
produce the dispersion relation of the linear system with the same potential
(\ref{Eq3}). The nonlinearity results in bifurcations of nonlinear modes
from different linear eigenstates, at respective eigenvalues. The spectrum
of the linear version of Eq.~(\ref{Eq4}) is composed of a series of discrete
eigenvalues [Figs.~\ref{fig1}(c,d)], supplemented by continuous spectrum
(not shown here), in contrast to the system with the HO potential, in which
case the discrete eigenvalues are distributed evenly.

The linear spectrum is determined by parameters of the trapping potential.
The variation of radius $r_{0}$, width $d$, and depth $p$ shift the spectrum
and alter the distribution of the eigenvalues in the spectrum on the other.
While fundamental QDs bifurcate from the eigenmode corresponding to the
first eigenvalue, dipole droplets bifurcate from the eigenmodes
corresponding to the second and third (mutually degenerate) eigenvalues. The
linear dipole modes corresponding to the second and third eigenvalues are
mutually perpendicular. For topological charge $m$, one can obtain droplets
with $2m$ poles from the eigenmodes corresponding to the eigenvalues with
numbers $n=2m$ and $2m+1$.

In addition to the fundamental and multipole droplets, VQDs can also
bifurcate from linear superpositions of two degenerate modes. For example,
VQDs with $m=\pm 1$ bifurcate from linear modes $\Psi _{\pm 1}=\Psi
_{1,2}\pm i\Psi _{1,3}$, where the first and second subscripts refer to the
topological charge and serial number of the basic eigenmodes, respectively.
More generally, VQDs with topological charge $m$ bifurcate from the
superposition of the pair of mutually degenerate linear modes $\Psi _{\pm
m}=\Psi _{m,2m}\pm i\Psi _{m,2m+1}$, which correspond to the same $2m$-th
eigenvalue.

\begin{figure}[tbph]
\centering
\includegraphics[width=0.45\textwidth]{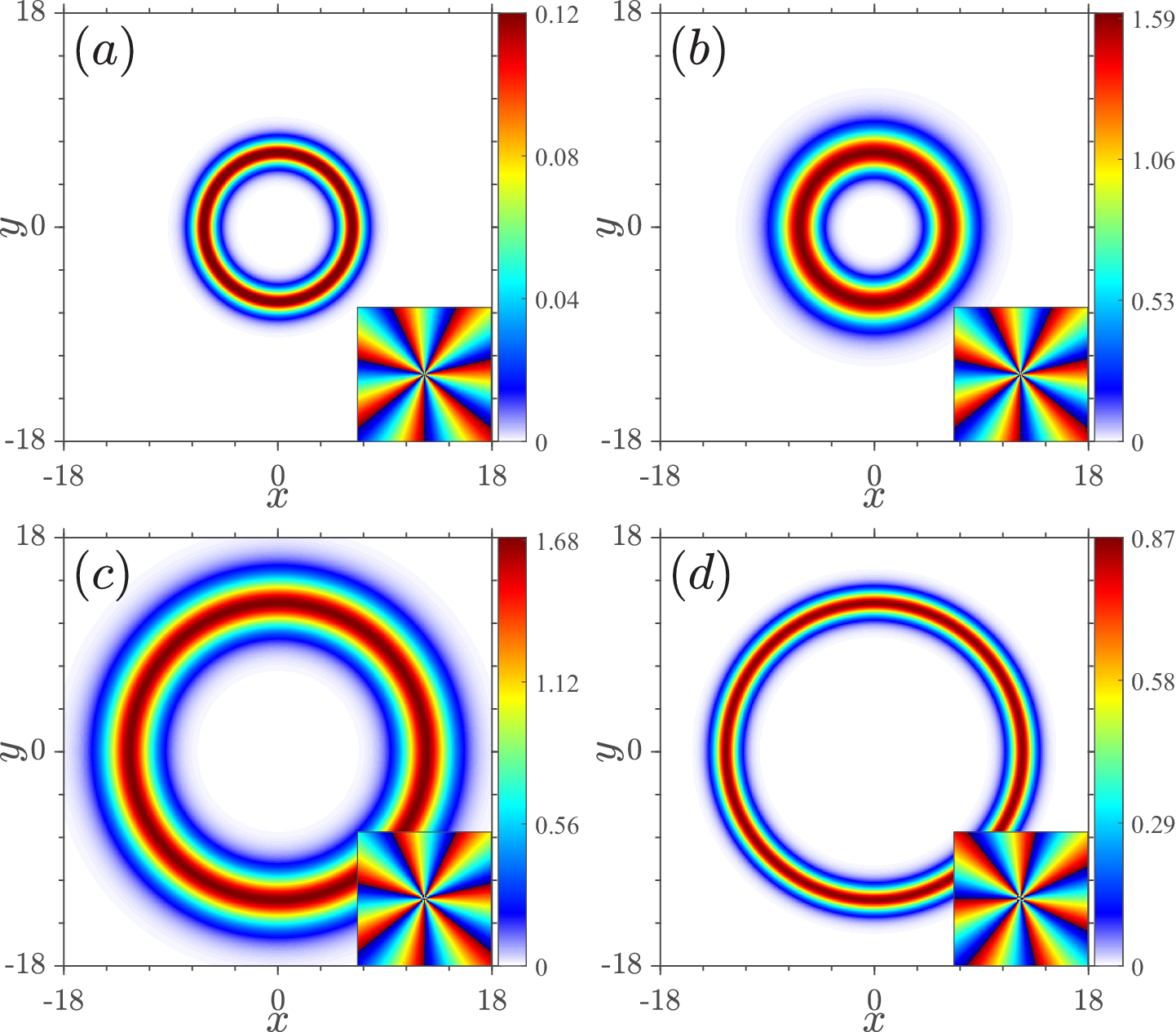}\vskip-0pc
\caption{Profiles of $\left\vert \protect\psi (r)\right\vert $ for VQDs with
$m=7$, marked in Fig.~\protect\ref{fig3}(a), in the annular potentials (%
\protect\ref{Eq3}) with $r_{0}=2\protect\pi ,d=1$ (a,b) and $r_{0}=4\protect%
\pi ,d=1$ (c,d). The chemical potential is $\protect\mu =-2.2$ in (a), $-0.5$
in (b,c), and $-2.9$ in (d). Insets: the corresponding phase structures. The
vortex shown in (a) belongs to the lower branch, and ones (b-d) belong to
the upper branch. In all the panels, the potential's depth is $p=4$. }
\label{fig2}
\end{figure}

It is known that, in the framework of Eq.~(\ref{Eq2}) in the free space ($V=0$), VQDs in the $2$D symmetric binary BEC can be stable with topological
charge up to $m=5$ (at least), provided that the atom number (norm) exceeds
a critical value \cite{PhysRevA.98.063602,PhysRevA.106.053303}. To
demonstrate the stabilization effect of annular potentials on VQDs with
higher charges, in Fig.~\ref{fig2} we produce several representative
examples of stable VQDs with $m=7$ in the potential (\ref{Eq3}) with $%
r_{0}=2\pi ,d=1$ and $r_{0}=4\pi ,d=1$ (recall $p=4$ is fixed). They exhibit
a ring-shaped profile of $\left\vert \psi (r)\right\vert $ distribution,
resembling the shape of the trapping potential, cf. Figs. \ref{fig1}(a,b).
The amplitude and thickness of the VQDs change with the variation of
chemical potential $\mu $. The radius at which $\left\vert \psi
(r)\right\vert $ attains its maximum is naturally constrained by radius $%
r_{0}$ of the annular potential. This fact indicates that VQDs are strongly
shaped by the trapping potential, in contrast to the flat-top profiles of
the free-space VQDs and vortex solitons in optical media with competing
nonlinearities, cf. Refs. \cite{PhysRevE.64.057601,PhysRevA.98.063602,PhysRevA.106.053303}.

\begin{figure}[t]
\centering
\includegraphics[width=0.48\textwidth]{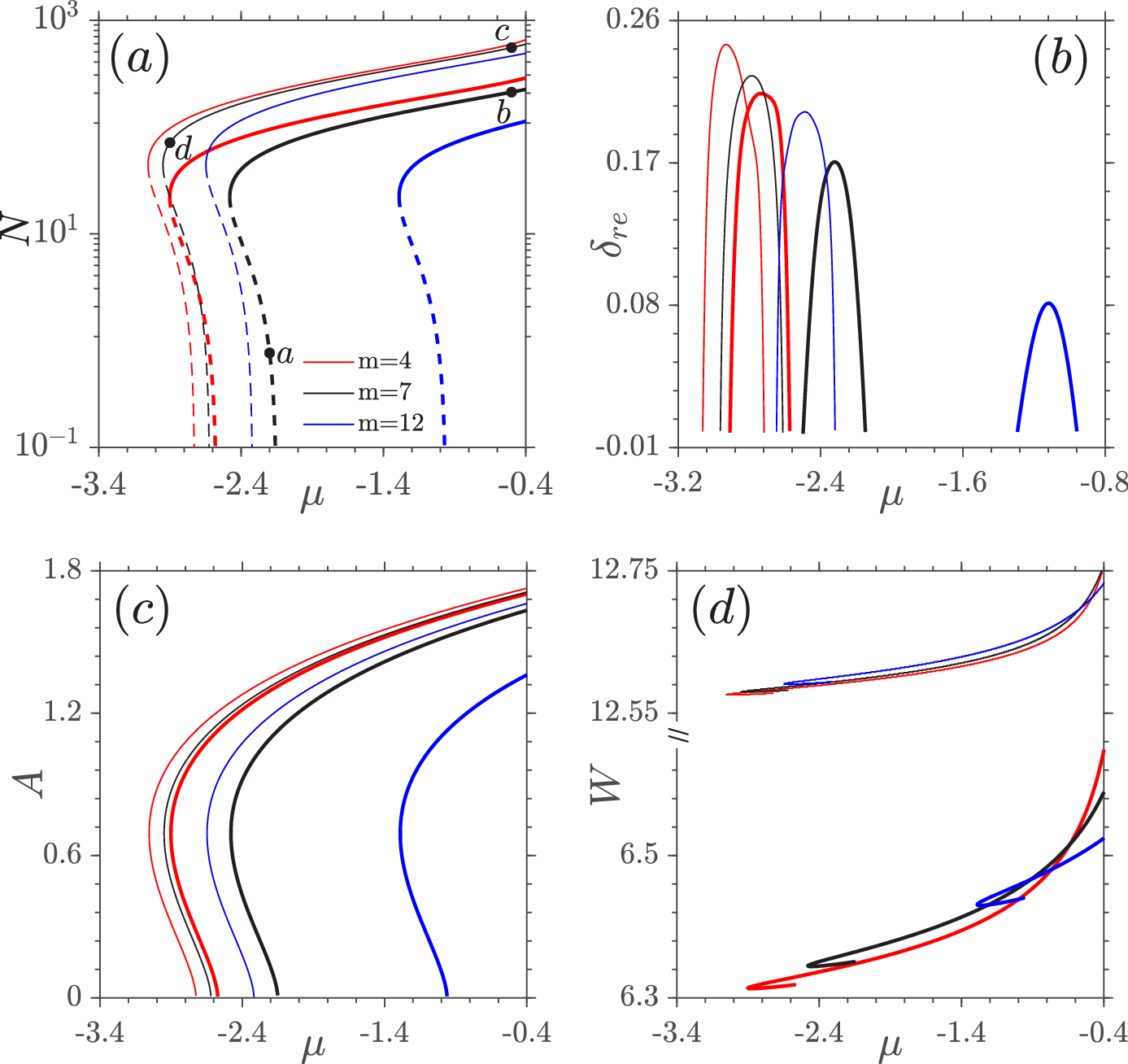}\vskip-0pc
\caption{(a) Norm $N$ vs. chemical potential $\protect\mu $ for VQDs with $%
m=4$, $7$ and $12$. Solid and dashed segments designate stable and unstable
states, respectively. (b) The instability growth rate $\protect\delta _{\text{%
re}}\equiv \mathrm{Re}(\protect\delta )$ vs. $\protect\mu $ for the VQD
solutions belonging to the lower branches of the $N(\protect\mu )$ families.
(c,d) The amplitude $A\equiv |\protect\psi |_{\text{max}}$ and width of the
VQDs with $m=4,7$ and $12$ versus $\protect\mu $. In all the panels, the
potential's depth is $p=4$. The data for the VQDs with $m=4,7$ and $12$ are
shown by red, black and blue curves, respectively. The thick and thin lines
correspond to the potential with $r_{0}=2\protect\pi $ and $4\protect\pi $.
In both cases, the potential's width is $d=1$. }
\label{fig3}
\end{figure}

Another essential difference from the free-space VQDs is the fact that the
presence of the annular potential makes the norm (scaled atom number) $N$ of
VQDs a nonmonotonous function of the chemical potential, as shown in Fig.~%
\ref{fig3}(a). First. $N$ increases first with the decrease of $\mu $, under
the action of the effective attractive nonlinearity in Eq.~(\ref{Eq4}) in
the case of $\left\vert \psi \right\vert <1$. The increase of $N$ is
accompanied by the increase of the droplet's amplitude, $|\psi |_{\text{max}%
} $, and its expansion, see Figs.~\ref{fig3}(c,d). In the case of $|\psi >1$%
, the logarithmic factor switches the sign of the nonlinearity to the
repulsion. The further increase of $N$ accelerates the growth rate of the
effective droplet's width, defined by relation
\begin{equation*}
W^{2}=N^{-1}\int \int (x^{2}+y^{2})\psi ^{2}\text{d}x\text{d}y,
\end{equation*}%
as seen in Fig.~\ref{fig3}(d), and slows down the growth rate of $|\psi |_{%
\text{max}}$ in Fig.~\ref{fig3}(c).

In the interplay with the trapping potential, the switch of the attractive
interaction into repulsive prevents the existence of the VQD branch with $%
dN/d\mu <0$ to the left of the cutoff point, i.e., at $\mu <\mu _{\text{cut}%
} $. As seen in Fig.~\ref{fig3}(a), at the cutoff point the terminating
lower VQD branch mergers with the upper one, with the same vorticity $m$ but
opposite sign of the slope, $dN/d\mu >0$. The upper branch is dominated by
the nonlinearity with the repulsive sign.

For the comparison's sake, Fig.~\ref{fig3}(a) also incudes the $N(\mu )$
dependences for the VQD families with $m=4$ and $12$ in the annular
potentials with $r_{0}=2\pi ,d=1$ and $r_{0}=4\pi ,d=1$ . The VQDs with
different values of $m$ bifurcate from combinations of linear mutually
degenerate states at the corresponding eigenvalues. Specifically, in the
potential with $r_{0}=2\pi ,d=1$ the VQDs with $m=4,7$ and $12$ bifurcate,
severally, from $\mu =-2.571,-2.149$, and $-0.960$ [these bifurcation points
are marked in Fig.~\ref{fig1}(c)]. The bifurcation values of $\mu $ for the
VQDs in the potential with $r_{0}=4\pi ,d=1$ are $-2.725$ (for $m=4$), $%
-2.619$ ($m=7$), and $-2.318$ ($m=12$). It is obvious that the $N(\mu )$
dependences are very similar for the VQDs with different vorticities $m$ in
the potentials with different widths $d$ are similar to each other. The
increase of the potential's radius shifts the norm curves leftward. The
value of the norm at the cutoff (branch-merger) point in the potential with
a larger radius ($r_{0}=4\pi $) is higher than in the case of the smaller
small radius ($r_{0}=2\pi $).

The central finding of this paper is that the annular potential added to the
LHY-amended GPE helps to stabilize VQDs with $m>5$. To illustrate this
point, we have conducted linear stability analysis of the obtained
stationary VQD solutions, based on BdGEs~(\ref{Eq7}). The smaller amplitude $%
|\psi |_{\text{max}}$ of the lower-branch VQD solutions makes the
self-attraction a dominant term, leading to the splitting instability of the
vortex modes, driven by the MI of the axisymmetric ring-shaped modes against
azimuthal perturbations. Indeed, the numerical solution of Eqs.~(\ref{Eq7})
yields nonzero instability growth rates for the lower-branch VQDs with
different vorticities $m\neq 0$, in the annular potentials (\ref{Eq3}) with
both $r_{0}=2\pi $ and $4\pi $ [Fig.~\ref{fig3}(b)].

Note that the negative slope of the lower branch $N(\mu )$ of the VQD
solutions, $dN/d\mu <0$, implies that this branch satisfies the well-known
Vakhitov-Kolokolov (VK) necessary stability condition, which actually
implies the absence of pure real unstable eigenvalues produced by the BdGEs
\cite{VK1973,BERGE1998259}. However, the splitting instability is not comprised by
the VK criterion, as it is accounted for by complex unstable eigenvalues
\cite{BERGE1998259,pego2002spectrally,malomed2022multidimensional}.

On the other hand, the numerical solutions of BdGEs (\ref{Eq7}) produces the
full spectrum of eigenvalues with Re$(\sigma)=0$ in their whole existence
domain, regardless of the values of potential's width, radius, and depth.
This result indicates that the upper-branch VQD families are completely
stable. As concerns the respective $N(\mu)$ dependences, which feature $%
dN/d\mu >0$ in Fig.~\ref{fig3}(a), they comply with the anti-VK stability
criterion \cite{PhysRevA.81.013624}, which is precisely $dN/d\mu >0$ in the case when bound
states are maintained by the self-repulsive nonlinearity.

We stress that the stable upper-branch VQD solutions feature a relatively
small number of atoms. In particular, the stability conditions for the
vortices with $m=7$ in the trapping potential (\ref{Eq3}) with $r_{0}=2\pi
,d=1$, or $r_{0}=4\pi ,d=1$ are, respectively, $N\geq 21.95$ or $N\geq 43.91$%
. This is in sharp contrast to the case without the trapping potential,
where the stability conditions for the VQDs with $m=2$ or $m=5$ are,
severally, $N\geq 200$ or $N\geq 3550$ \cite{PhysRevA.98.063602}. The VQD
norm in the free space dramatically increases, featuring $N\rightarrow
\infty $, as $\mu $ is approaching the upper cutoff. For instance, the
chemical potential of the $m=5$ VQD with $N=3550$ is $\mu =-0.297$, which is
close to the respective cutoff point, $\mu _{\text{cut}}=-0.304$ \cite%
{PhysRevA.106.053303}. Thus, the addition of the annular trapping potential
to the LHY-amended BEC model drastically reduces the number of atoms
required for the creation of stable higher-order VQDs, making it much easier
to realize the predictions in the experiments.

\begin{figure}[tbph]
\centering
\includegraphics[width=0.48\textwidth]{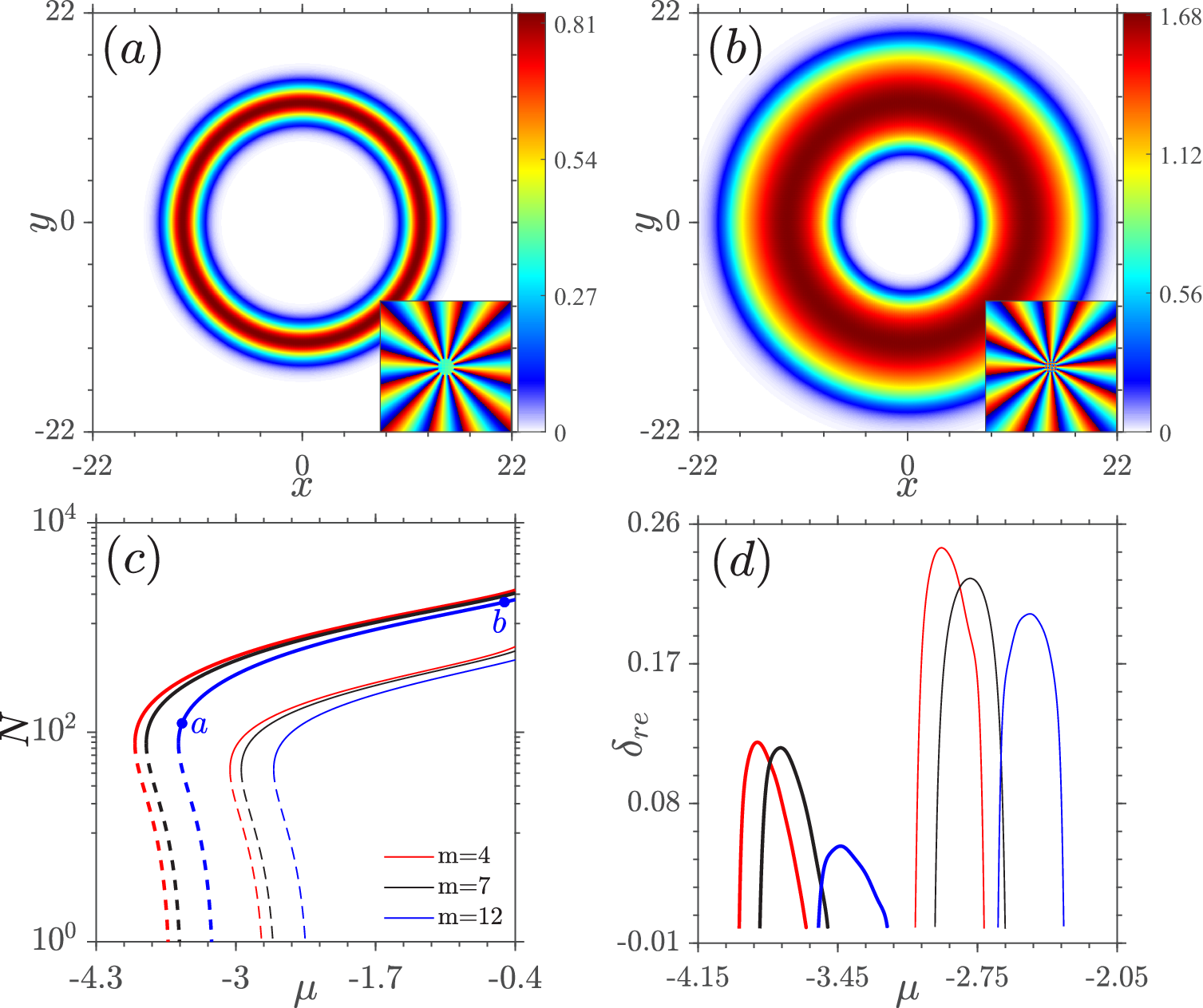}
\caption{Examples of the upper-branch VQDs with $m=12$ marked in (c) at $%
\protect\mu =-3.5$ (a) and $-0.5$ (b), in the trapping potential (\protect
\ref{Eq3}) with $r_{0}=4\protect\pi $ and $d=4$. (c) Norm $N$ vs. $\protect%
\mu $ for the VQDs with $m=4$, $7$ and $12$. Solid and dashed curves depict
stable and unstable subfamilies, respectively. (d) The stability growth rate
$\protect\delta _{\text{re}}\equiv \mathrm{Re}(\protect\delta )$ vs. $%
\protect\mu $ for the lower-branch VQDs (the red, black, and blue branches
represent $m=4$, $7$, and $12$, respectively). The thick and thin lines
correspond to the annular potential with $r_{0}=4\protect\pi ,d=4$ and $%
r_{0}=4\protect\pi ,d=1$, respectively. In all the cases, the potential's
depth is $p=4$. }
\label{fig4}
\end{figure}

\begin{figure*}[h]
\centering
\includegraphics[width=0.85\textwidth]{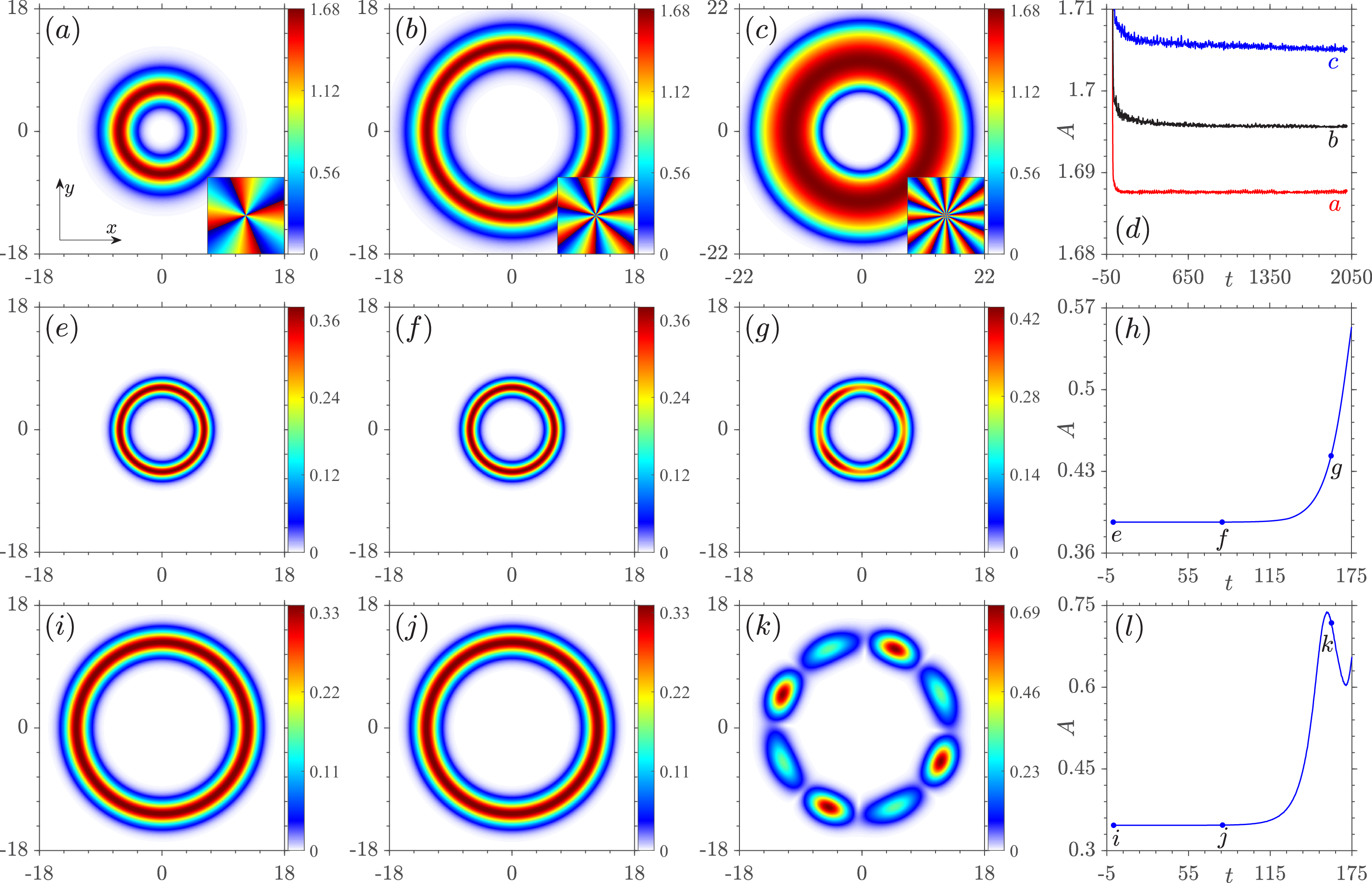}
\caption{(a-c) Vortex profiles with $m=4, 7$ and $12$ marked in (d) at $t=2000$. The potential
parameters $r_0=2\protect\pi, d=1$ in (a), $r_0=4\protect\pi, d=1$ in (b)
and $r_0=4\protect\pi, d=4$ in (c). Insets: the corresponding phase
patterns. (d) Dependence of peak amplitudes $A=\text{max}|\Psi(t)|$ of stable
upper-branch vortex droplets on $t$. The blue curve is increased by 0.01 to avoid the overlaps with the black curve. The chemical potential $\protect\mu=-0.5$ in (a-d). (e-f) Unstable
evolution of the lower-branch fundamental droplet marked in (h) at $\protect%
\mu=-3$ in the potential with $r_0=2\protect\pi, d=1$. (h) Peak amplitude $%
\text{max}|\Psi(t)|$ of the unstable fundamental droplet shown in (e) versus
$t$. (i-k) Unstable evolution of the lower-branch VQD with $m=12$ marked in (l) at $\protect \mu=-3.4$ in the potential with $r_0=4\protect\pi, d=4$. (h) Peak amplitude $\text{max}|\Psi(t)|$ of the unstable vortex shown in (i) versus $t$. }
\label{fig5}
\end{figure*}

The dependence of the amplitude, $|\psi |_{\max }$, of the VQDs with
different vorticities, trapped in the annular potentials with different
thicknesses, on the chemical potential $\mu $ is shown in Fig.~\ref{fig3}(c). At first, the amplitude increases with the decrease of $\mu $. Passing
the turning (cutoff) point, it keeps increases with the subsequent growth of
$\mu $, although slower. At fixed $\mu $, the effective width of the VQDs in
the trapping potential with $r_{0}=4\pi $ is approximately twice that in the
potential with $r_{0}=2\pi $, see Fig.~\ref{fig3}(d). This finding again
confirms that the VQDs are confined within the ring potential. Adjusting the
potential's radius, one can create VQDs with any desired radius.

To study the stability of VQDs with high topological charges, we consider
them with $m=1$ to $12$ in setups with different parameters of the annular
potential (\ref{Eq3}). For illustration, in Figs.~\ref{fig4}(a,b) we present
the profiles of $|\psi |$ of two VQDs with $m=12$ at different values of $%
\mu $ in the annular potential with $r_{0}=4\pi $ and $d=4$. The vortex is
thin at large $|\mu |$ and thick at small $|\mu |$. Due to the strong
repulsive nonlinearity and the confinement imposed by the annular potential,
with the increase of $\mu $, the VQD demonstrates symmetric expansion with
respect to the central circle (where the potential attains its maximum). The
dependence $N(\mu )$ for VQDs with vorticities $m=4,7$ and $12$ is shown in
Fig.~\ref{fig4}(c). The bifurcation points at which the VQDs emerge from the
linear modes are $\mu =-3.607$, $-3.502$, and $-3.202$ for $m=4$, $7$ and $%
12 $, respectively. The norm remains a nonmonotonous function of $\mu $,
similar to the $N(\mu )$ curves shown in Fig.~\ref{fig3}(a). The growth of
the potential's width $d$ leads to an increase of the critical values of $N$
at the turning points, where the lower and upper branches merge. The growth
of $d$ also shifts the $N(\mu )$ curve leftward.

The stability analysis results shown in Fig.~\ref{fig4}(d) demonstrate that,
as said above, all VQDs belonging to the lower branches (with $dN/d\mu <0$)
are unstable. Yet, the instability growth rate for these solutions decreases
with the increase of width $d$ of the annular trapping potential (\ref{Eq3}). As a result, one can obtain nearly stable VQDs belonging to the lower
branch, with small atom numbers.

The complete stability region of the upper-branch VQDs is in sharp contrast
to that for the upper-branch VQDs trapped in the HO potential \cite%
{LIU2023113422}, where the stability region quickly shrinks with the growth
of the vorticity.

Although we have displayed in detail the examples of VQDs with $m=4,7$ and $%
12$, our main results hold as well for all VQDs with even and odd vorticity,
at least for $m\leq 12$. The properties of VQDs also remain qualitative unchanged for varying potential depths $p$. It is plausible that the upper-branch VQDs remain
stable for even higher values of $m$. We have also checked the stability of
the fundamental QDs with $m=0$ in the annular potentials, concluding that
the fundamental QDs belonging to the corresponding lower branch remain
unstable for all values of the parameters of the annular trapping potential.
As well as the lower-branch vortex modes, this is the azimuthal MI, as
confirmed by direct simulations below, see Figs.~\ref{fig5}(e-l).

To validate the predictions of the linear-stability analysis, we have
perform extensive evolution simulations of VQDs with different topological
charges in the trapping potential with various parameters, by means of the
split-step Fourier method with absorptive boundary condition. The perturbation was added, as white noise in the form $\Psi (x,y,t=0)=\psi (x,y)[1+\rho (x,y)]$, to the input at $t=0$ for the stable VQDs and no noise was added to the unstable droplets, whose instability was predicted by the computation of eigenvalues for small perturbations. For unstable VQDs, the random perturbation was taken with variance $\sigma _{\text{noise}}=0.01$.

Typical examples of the simulated perturbed solutions are displayed in Fig.~\ref{fig5}. In particular, the inputs used in panels (b) and (c) correspond
to the stationary VQDs shown in Figs.~\ref{fig2}(b) and \ref{fig4}(b),
respectively. The upper-branch VQDs which are predicted to be stable indeed
preserve their amplitude and phase structures in the course of arbitrarily
long simulations, as shown in Figs.~\ref{fig5}(a-d). On the other hand, the
unstable lower-branch VQDs spontaneously break up into one or several
azimuthal fragments after a short evolution time. To further highlight the
instability of the lower-branch VQDs, we present examples for the
fundamental droplet (with $m=0$) in Figs.~\ref{fig5}(e-h) and for the VQD ($m=12$) in Figs.~\ref{fig5}(i-l). As previously stated, the direct simulations clearly demonstrate that the onset of the azimuthal MI occurs.

\section{Conclusion}

\label{Sec4} In this work we have investigated the stability and dynamics of
VQDs (vortex quantum droplets) in the 2D symmetric binary BEC trapped in the
annular potential. Two branches of the VQD solutions, with opposite slopes
of the norm-vs.-chemical-potential curves, are supported by the mean-field
nonlinearity amended by the LHY (Lee-Hung-Yang) correction, i.e., the cubic
term times the logarithmic factor. Although the profiles of VQDs vary with
the chemical potential and topological charge $m$, they are confined by the
trapping potential, providing a new way to control their radius, width, and
amplitude. While the lower-branch VQDs, bifurcating from the linear modes,
are unstable in their entire existence domain, the upper branch is
completely stable, for all values of $m$ and potential's parameters. The
stability of the upper-branch VQDs with $m\leq 12$ (at least), for which the
effective nonlinearity is self-repulsive, agrees with the \textit{anti-Vakhitov-Kolokolov} criterion. We have also produced analytical results for the MI (modulational instability) of quasi-1D VQDs, trapped in an extremely tight annular potential. The findings propose an effective way for the creation of stable higher-order vortex droplets with a relatively small
number of atoms, which is favorable for the experimental realization.




\vskip0.5pc \textbf{CRediT authorship contribution statement}

Liangwei Dong: Conceptualization and writing the original draft; Mingjing
Fan: Numerical calculations; Boris A. Malomed: Analytical investigation,
writing, review $\&$ editing and validation.

\vskip0.5pc \textbf{Declaration of competing interest} The authors declare
the following financial interests/personal relationships which may be
considered as potential competing interests: Liangwei Dong reports financial
support provided by Natural Science Basic Research Program in Shaanxi
Province of China (grant No. 2022JZ-02); Boris A. Malomed reports financial support provided by
the Israel Science Foundation through grant No. 1695/22.

\vskip0.5pc \textbf{Data availability} Data will be made available on request.

\vskip0.5pc \textbf{Acknowledgements} This work is supported by the Natural
Science Basic Research Program of Shaanxi Province of China (Grant No.
2022JZ-02) and Israel Science Foundation (Grant No. 1695/22).



\end{document}